\documentclass[twocolumn,fp,seceq]{jpsj3}
\newcommand{\ket}[1]{\left\vert {#1} \right\rangle}

\def\Jeff{J_{\rm eff}}
\def\Deff{D_{\rm eff}}
\def\tJeff{{\tilde{J}}_{\rm eff}}
\def\tDeff{{\tilde{D}}_{\rm eff}}
\def\De{\tilde{D}}
\def\Do{D}
\def\Nc{N_{\rm c}}
\def\dfrac#1#2{{\displaystyle\frac{#1}{#2}}}
\def\szcl#1{{S^z_{\rm tot}({#1})}}

\def\Heff{{\cal{H}}_{\rm eff}}
\def\eff{{\rm {eff}}}
\def\ln{{\rm{ln}}}

\def\v#1{\mib #1}

\title
{Ground-State Phases of Anisotropic Mixed Diamond Chains with Spins $1$ and $1/2$}

\author
{Kazuo Hida\thanks{E-mail: hida@mail.saitama-u.ac.jp} 
}

\inst
{Division of Material Science, Graduate School of Science and Engineering, \\ Saitama University, Saitama, Saitama 338-8570, Japan
}

\recdate
{August 18, 2014
}

\abst{
The ground-state phases of anisotropic mixed diamond chains with spins 1 and 1/2 are investigated. Both  single-site and exchange anisotropies are considered. We find the phases 
consisting of  an array of uncorrelated spin-1 clusters separated by singlet dimers. Except in the simplest case where the cluster consists of a single $S=1$ spin, this type of ground state breaks the translational symmetry spontaneously. Although the mechanism leading to this type of ground state is the same as that in the isotropic case, it is nonmagnetic or paramagnetic depending on the competition between two types of anisotropy. 
We also find the N\'eel, period-doubled N\'eel, Haldane, and large-$D$ phases, 
where the ground state is a single spin cluster of infinite size equivalent to the spin-1 Heisenberg chain with alternating anisotropies. The ground-state phase diagrams are determined for typical sets of parameters by  numerical analysis.  In various limiting cases, the ground-state phase diagrams are determined analytically. The low-temperature behaviors of magnetic susceptibility and entropy are investigated to distinguish each phase by observable quantities. The relationship of the present model with the anisotropic rung-alternating ladder with spin-1/2 is also discussed.
}

\begin{document}

\sloppy
\maketitle
\section{Introduction}

In recent decades, the interplay of frustration and quantum fluctuation in spin systems has motivated a number of theoretical and experimental investigations.\cite{diep,intfrust} Among them, in the theoretical approach, exactly solvable models such as the Majumdar-Ghosh\cite{mg} and Shastry-Sutherland\cite{shs} models have played  crucial roles in the understanding of the nature of frustrated quantum magnetism. 

\begin{figure} 
\centerline{\includegraphics[width=4.5cm]{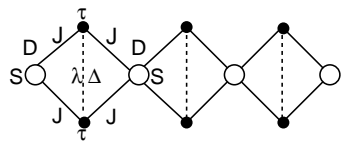}}
\caption{
Structure of the diamond chain. Spin magnitudes in a unit cell are indicated by $S$ (open circles) and $\tau$ (filled circles) 
 The parameters $J$, $\lambda$, $D$, and $\Delta$ in the Hamiltonian (\ref{hama}) associated with each site and bond are indicated.

}
\label{lattice_structure}
\end{figure}

In these two models, the ground states  are dimer states. In contrast, the diamond chain is a frustrated spin chain with exact spin-cluster-solid ground states that are different from dimer states.  The lattice structure of the diamond chain is shown in Fig.~\ref{lattice_structure}. In a unit cell, there are two nonequivalent sites occupied by spins with magnitudes $S$ and $\tau$; we denote the set of magnitudes by ($S$, $\tau$). Takano and coworkers\cite{takano,Takano-K-S} introduced this lattice structure and  generally investigated the case of ($S$, $S$). They found that several 
 states that consist of a regular array of spin clusters are the exact ground states of this model. Except in the simplest case where the cluster consists of a single spin, this type of ground state breaks the translational symmetry spontaneously. 
 Later,  Niggemann et al.\cite{nig1,nig2} investigated a series of diamond chains with ($S$, 1/2).\cite{tsh} Recently, extensive investigation on the mixed diamond chains (MDCs) with ($S$, $S/2$) for the integer $S$ has been carried out by the present author and coworkers\cite{tsh,hts,htsalt,htsus} mainly for $S=1$. The MDCs are of special interest among diamond chains, because only the MDCs have the nonmagnetic ground state in the absence of  frustration owing to the Lieb-Mattis theorem. Hence, the frustration-induced transitions from various nonmagnetic phases to spin-cluster-solid phases can take place in the MDC. 

Diamond chains have an infinite number of local conservation laws, and more than two different types of exact  
ground state are realized depending on the strength of frustration. 
 Hence, using diamond chains, we can investigate various aspects of frustrated magnetism on a well-founded basis. This motivated the investigation of various modifications of the diamond chain by  many researchers.\cite{ottk,otk,sano,dia4spin,htanis,htsalt,htsus} 
 In addition, as reviewed in Ref. \citen{htsalt}, the MDC is related to many other important models of frustrated magnetism.

In the present work, we investigate  the ground-state properties of the MDC with (1,1/2) that has the single-site anisotropy $D$ on the $S=1$ site and exchange anisotropy $\Delta$ on the bond connecting two $S=1/2$ sites. In the following, we call this model  an anisotropic MDC (AMDC). The anisotropy $\Delta$ plays the role equivalent to the single-site anisotropy, if the pair of $S=1/2$ spins forms a $S=1$ triplet state.  
The alternating $D$-terms in  spin-1 Heisenberg chains can pin the fluctuating hidden order to induce a long-period N\'eel order.\cite{hc}
 Considering the high degeneracy of the ground states of the MDC,\cite{tsh} we may expect an even richer phase diagram in the AMDC. In the presence of anisotropy, various limiting cases with strong anisotropies can be analytically investigated. In these limits, the spin configurations of the ground states are explicitly determined, and the physical picture of the ground states and the transitions between them can be clarified.

Despite the theoretical relevance of the MDC, no materials described by the MDC have been found so far. Nevertheless, synthesizing MDC materials is a realistic expectation in view of the successful  synthesis of many novel magnetic materials such as molecule-based magnetic materials.\cite{m-d}
 From this viewpoint, it is important to present theoretical predictions on the ground-state properties of the AMDC to widen the range of candidate materials of the MDC and raise the possibility of their synthesis. For this purpose, we present the low-temperature behaviors of magnetic susceptibility and entropy in each phase and phase boundary.

This paper is organized as follows. 
In sect. 2, the Hamiltonian for the MDC with the single-site and exchange anisotropies is presented. 
In sect. 3, the ground-state phase diagram is obtained by the numerical method. 
Various limiting cases are also discussed. In sect. 4, the low-temperature behaviors of magnetic susceptibility and entropy are investigated. In sect. 5, the relationship between the AMDC and the anisotropic rung-alternating ladder (ARAL)\cite{tone_prv} is discussed. The last  section is devoted to the summary and discussion.

\section{Hamiltonian}
We consider the AMDC with $(S, \tau)=(1,1/2)$ described by the   Hamiltonian
\begin{align}
{\cal H} &=\sum_{l=1}^{L} \Big[J(\v{S}_{l}+\v{S}_{l+1})\cdot(\v{\tau}^{(1)}_{l}+\v{\tau}^{(2)}_{l})+{\Do}S^{z2}_l\nonumber\\
&+\lambda\left\{\Delta{\tau}^{z(1)}_{l}{\tau}^{z(2)}_{l}+ {\tau}^{(1)x}_{l}{\tau}^{(2)x}_{l}+ {\tau}^{(1)y}_{l}{\tau}^{(2)y}_{l}\right\}\Big],
\label{hama}
\end{align}
where $\v{S}_{l}$ is the spin-1 operator, and 
$\v{\tau}^{(1)}_{l}$ and $\v{\tau}^{(2)}_{l}$ 
are the spin-1/2 operators in the $l$th unit cell. The total number of unit cells is denoted by $L$.  We consider the case of $\lambda > 0$ unless specifically mentioned.

The Hamiltonian (\ref{hama}) has a series of conservation laws. This can be seen by rewriting  Eq. (\ref{hama})  in the form 
\begin{align}
{\cal H} &=\sum_{l=1}^{L} \left[J(\v{S}_{l}+\v{S}_{l+1})\cdot\v{T}_{l}+{\Do}S^{z2}_l
+{\De}{T}^{z 2}_{l}
+\frac{\lambda}{2}\v{T}_{l}^2\right]\nonumber\\
&-\frac{\lambda L}{4}(\Delta+2),
\label{ham2}
\end{align}
where 
\begin{align}
{\De}=\frac{\lambda}{2}(\Delta-1)\label{eq:defde}
\end{align}
using the relations $\v{\tau}_l^2=3/4$ and  ${\tau}_l^{z2}=1/4$. Note that ${\tau}_l^{z2}$ is a c-number only for $\tau=1/2$. Hence, our method only applies to the AMDC with $\tau=1/2$. The composite spin operator $\v{T}_l$ is defined as 
\begin{align}
\v{T}_{l} \equiv \v{\tau}^{(1)}_{l}+\v{\tau}^{(2)}_{l} 
\quad (l = 1, 2, \cdots L). 
\end{align}
Then, it is  evident that 
\begin{align}
[\v{T}_l^2, {\mathcal H}] = 0 \quad (l = 1, 2, \cdots L). 
\end{align}
Thus, we have the $L$ conserved quantities $\v{T}_l^2$ for all $l$, even for $\Do \neq 0$ and $\Delta\neq 1$. 
Defining the magnitude $T_l$ of  $\v{T}_l$ by $\v{T}_l^2 = T_l (T_l + 1)$, we have  a 
 set of good quantum numbers $\{T_l; l=1,2,...,L\}$. 
Each $T_l$ takes a value of 0 or 1. 
The total Hilbert space of the Hamiltonian (\ref{ham2}) consists of 
separate subspaces, each of which is specified by 
a definite set of $\{T_l\}$, i.e., a sequence of 0 and 1.

A spin pair of $T_l=0$ is a singlet dimer that cuts off the correlation between $\v{S}_{l}$'s at  
both sides as seen 
from Eq.~(\ref{ham2}). 
Hence, when a segment is bounded by two $T_l=0$ pairs, it is isolated from 
 other parts of the spin chain. 
The segment including $n$ successive $\v{T}_{l}$'s  with $T_l=1$ and $n+1$ $\v{S}_{l}$'s 
 is called a cluster-$n$ as in the isotropic case.\cite{tsh} 
A  cluster-$n$ is equivalent to an antiferromagnetic Heisenberg chain consisting  of $2n+1$ effective spins with magnitude 1 
 with alternating anisotropy. 
The  Hamiltonian of  cluster-$n$ is given by
\begin{align}
{\mathcal {H}}_n = \sum_{l=1}^{n}J({\v{S}}_{l}+{\v{S}}_{l+1}){\v{T}}_{l}+\sum_{l=1}^{n+1}{\Do}{S}^{z2}_{l}
+\sum_{l=1}^{n}{\De}{T}^{z 2}_{l}.
\label{ham3}
\end{align}

The ground state of the AMDC is analyzed by describing each cluster in terms of 
 the finite-length spin-1 chain (\ref{ham3}). 
As in the isotropic case,\cite{htanis} we find successive phase transitions between dimer-cluster-$n$ (DC$n$) phases with $n \geq 0$ and $\infty$ as $\lambda$ decreases.  The DC$n$ phase is the phase where the ground state  consists of an alternating array of cluster-$n$'s and singlet dimers.   
The DC$\infty$ phase is equivalent to the ground state of the Hamiltonian (\ref{ham3}) with infinite length with $T_l = 1$ for all $l$.  
We follow Refs. \citen{tsh,hts,htsalt,htsus,htanis} to determine the phase boundary between  the DC$n$ phases with different $n$ values. 
In the DC$n$ phase, the ground-state energy $E_{\rm G}^{\rm tot}(L)$ of an entire chain with macroscopic length $L$ is given by
\begin{align}
E_{\rm G}^{\rm tot}(L)&=\frac{L}{n+1}\left(E_{\rm G}(2n+1,{\Do},{\De})-\lambda\right)-\frac{\lambda L}{4}(\Delta-2),
\end{align}
where $E_{\rm G}(2n+1,{\Do},{\De})$ is the ground-state energy of the Hamiltonian (\ref{ham3}), which can be easily obtained up to $n=7$ by the Lanczos and Householder numerical diagonalization method.

If a direct transition occurs between the DC$n_1$ and DC$n_2$ phases, the critical point is determined by equating the corresponding ground-state energies as
\begin{align}
\lambda_{\rm c}(n_1,n_2)&=\frac{1}{n_1-n_2}((n_1+1)E_{\rm G}(2n_2+1,{\Do},{\De})\nonumber\\
&-(n_2+1)E_{\rm G}(2n_1+1,{\Do},{\De})).\label{eq:lamcgen}
\end{align}
Hence, the DC$n$-DC$(n-1)$ critical point is given by
\begin{align}
\lambda_{\rm c}(n,n-1)&=(n+1)E_{\rm G}(2n-1,{\Do},{\De})\nonumber\\
&-nE_{\rm G}(2n+1,{\Do},{\De}).\label{eq:lamcnnp}
\end{align}
If we set $n_1\rightarrow\infty$ and $n_2=n$, we find the DC$n$-DC$\infty$ critical point as
\begin{align}
\lambda_{\rm c}(\infty,n)&=E_{\rm G}(2n+1,{\Do},{\De})-(n+1)\epsilon_{\rm G}({\Do},{\De}),\label{eq:lamcninf}
\end{align}
where $\epsilon_{\rm G}(D,\tilde{D})$ is the ground-state energy per unit cell of the $S=1$ chain (\ref{ham3}) with infinite length. We determine this value by the infinite-size DMRG method.

\section{Ground-State Phase Diagram}
\subsection{Numerical results}
\begin{figure}
\centerline{\includegraphics[width=5cm]{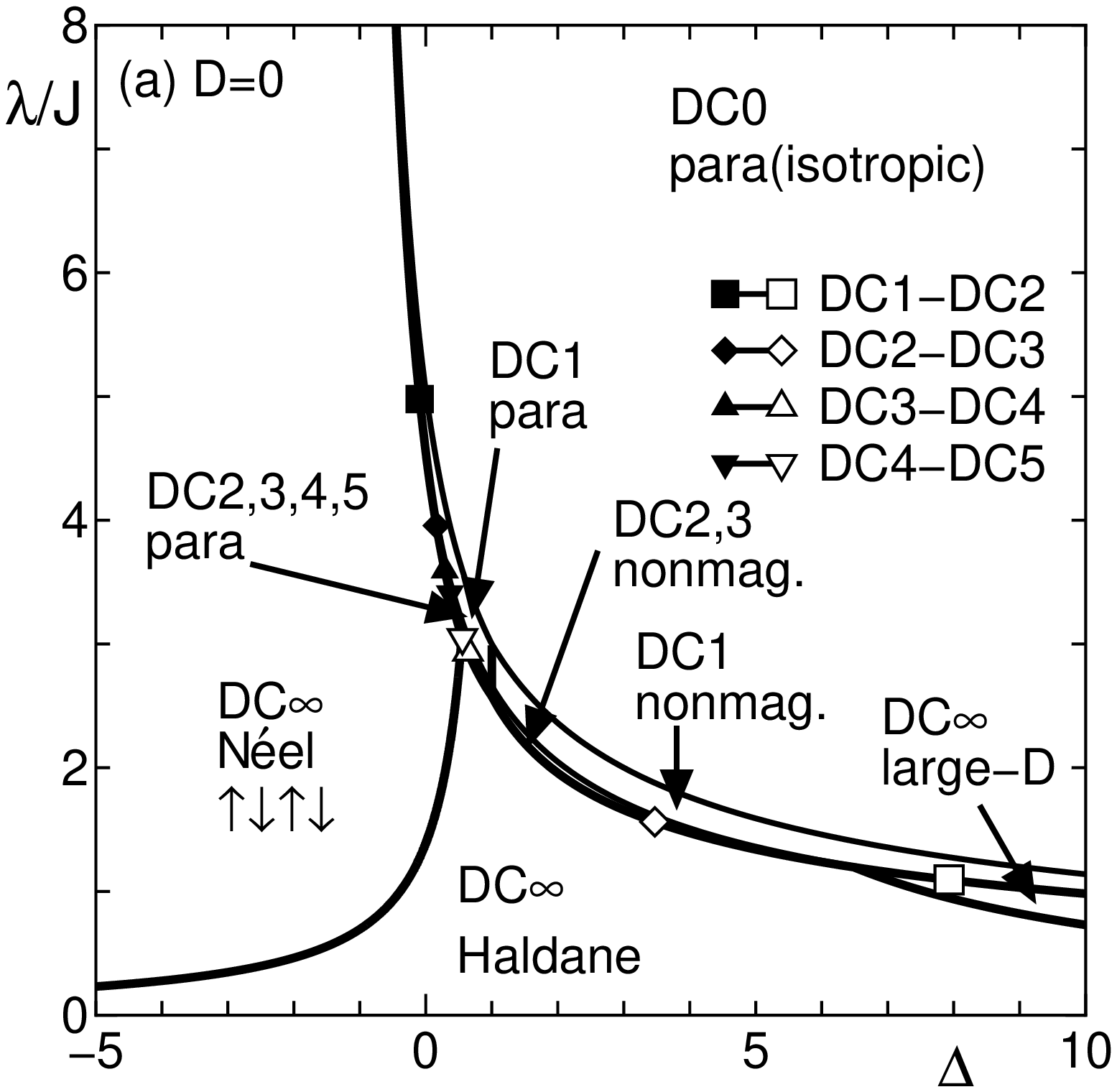}}
\centerline{\includegraphics[width=5cm]{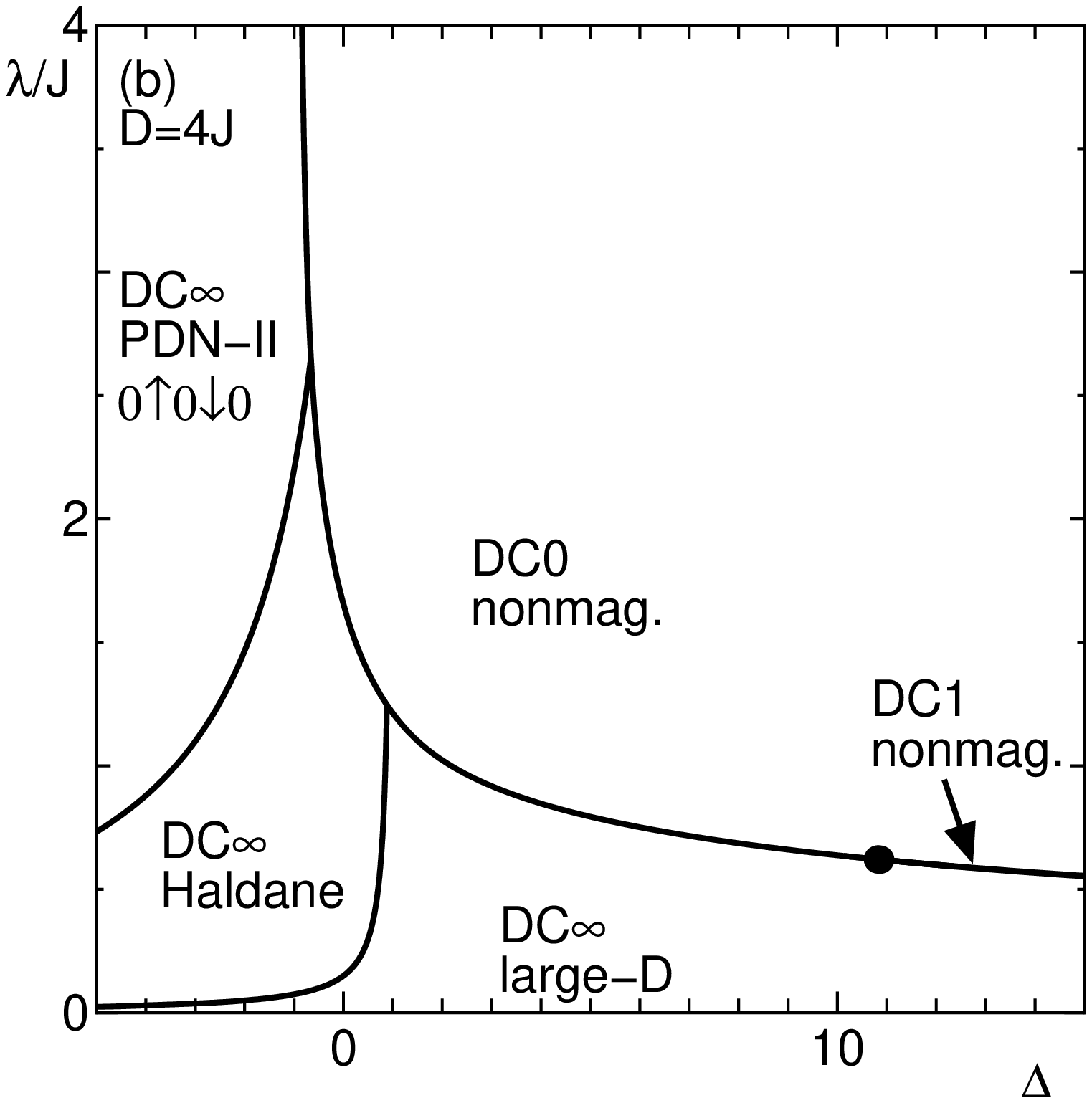}}
\centerline{\includegraphics[width=5cm]{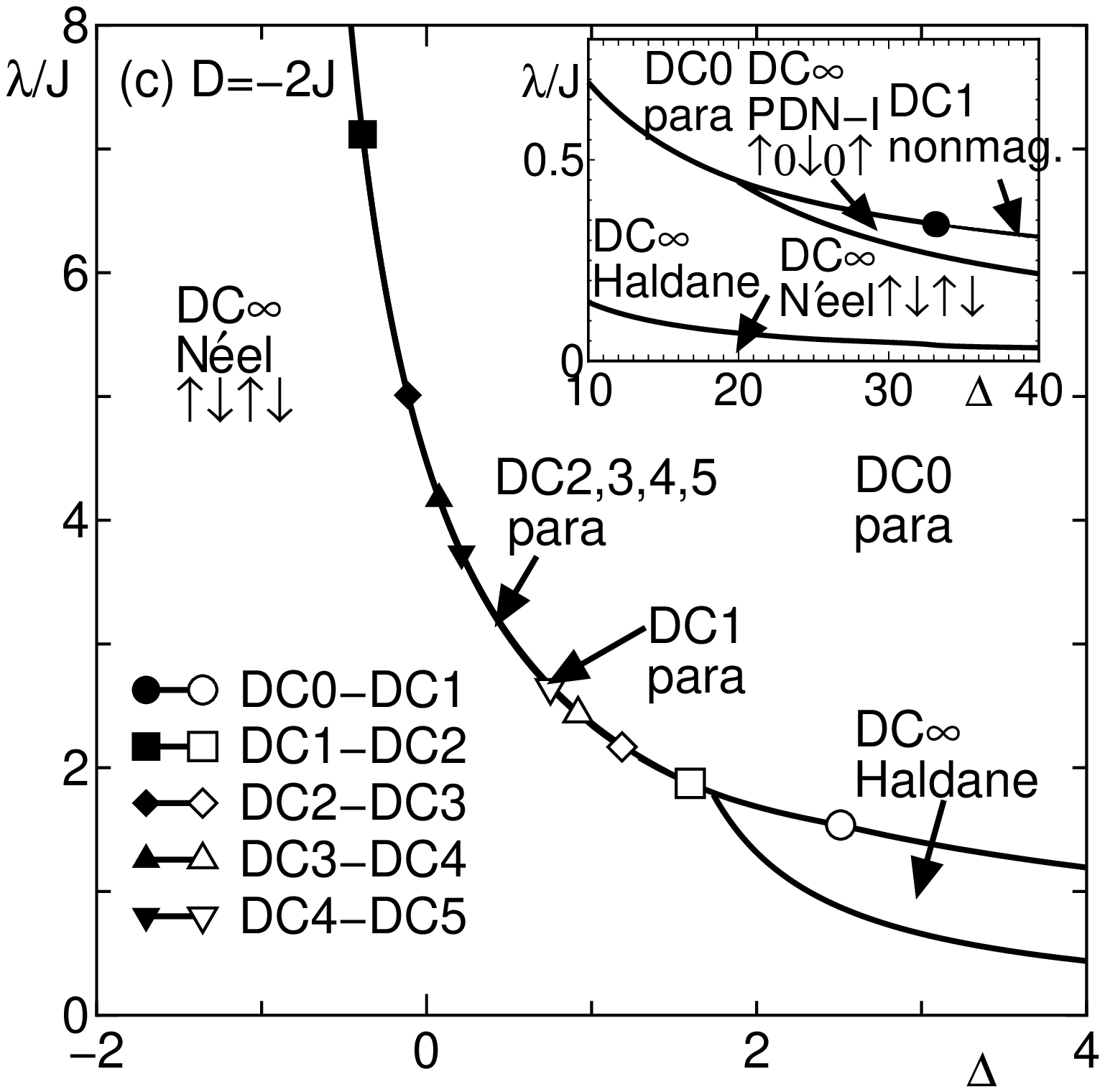}}
\caption{ Phase diagrams for (a) ${\Do}=0$, (b) ${\Do}=4J$, and  (c) ${\Do}=-2J$. The DC$n$ phases with $n \geq 1$ are almost invisibly narrow on the present scale except for the DC1 phase with $D=0$. The open (filled) symbols are their right (left) end points. In (c), the DC1 phase vanishes for $2.5 \lesssim \Delta \lesssim 33.2$.}
\label{fig:do00}
\end{figure}
\begin{figure}
\centerline{\includegraphics[width=5cm]{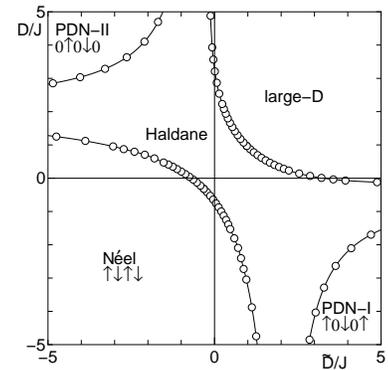}}
\caption{ Phase diagram  of a spin-1 Heisenberg chain with alternating anisotropies $\Do$ and ${\De}$ based on Ref. \citen{hc}. }
\label{fig:phasedual}
\end{figure}
The numerically obtained phase diagrams on the $\Delta-\lambda$ plane are shown in Fig. \ref{fig:do00} for $D=0$, $4J$, and $-2J$. In general, the ground state has the following properties:
\begin{enumerate}
\item The ground state of cluster-$n$ has $\szcl{n}=0$ or $\szcl{n}=\pm 1$ depending on the anisotropies $D$ and $\Delta$. Here, $\szcl{n}$ is the $z$-component of the total spin in the ground state of cluster-$n$. In the case of $\szcl{n}=0$, the DC$n$ phase is nonmagnetic, and in the case of $\szcl{n}=\pm 1$, it is paramagnetic. Except in the case of the cluster-0 with $\Do=0$, $\szcl{n}$ is  numerically determined for each cluster-$n$. 
The DC$n$ phases with a large $n$ are extremely narrow even if they exist. 
\item The DC$\infty$ phase is the ground state of a spin-1 Heisenberg chain with the alternating anisotropies $\Do$ and ${\De}$ described by the Hamiltonian (\ref{ham2}) with $T_l=1$ for all $l$.  The classification of the DC$\infty$ phases is given on the basis of the phase diagram of the spin-1 Heisenberg chain with alternating single-ion anisotropy\cite{hc} where $D_0+\delta D$ and $D_0-\delta D$ in Ref. \citen{hc} correspond to $\Do$ and $\De$, respectively.  
The conventional N\'eel, period-doubled N\'eel (PDN), Haldane, and large-$D$ phases can be realized depending on $\Do$ and $\De$, as shown in Fig. \ref{fig:phasedual}. There exist two types of PDN phase. In the PDN-I phase, the N\'eel order is supported by the spins $\v{S}_l$, while in the PDN-II phase, it is supported by the spins $\v{T}_l$.\cite{hc} 

\end{enumerate}
Particularly for $\Do=0$, the ground state has the following properties:
\begin{enumerate}
\item The ground state of cluster-$0$ is threefold-degenerate with $\szcl{0}=0, \pm 1$. Hence, the DC0 phase is always paramagnetic and isotropic.
\item The ground state of cluster-$n$ with $n \geq 1$ is a singlet with  
$\szcl{n}=0$ for $\Delta>1$, whereas it is a doublet with 
 $\szcl{n}=\pm 1$ for $\Delta < 1$. Hence, the DC$n$ phase with $\Delta >1$ is nonmagnetic and that with $\Delta \leq 1$ is paramagnetic.
\end{enumerate}

\subsection{Limiting case of $\Delta \rightarrow \infty$}
\subsubsection{Effective Hamiltonian}
In the limiting case of $\Delta\rightarrow \infty$, $\De$ also tends to $\infty$ from Eq. (\ref{eq:defde}). Hence, we express the Hamiltonian (\ref{ham3}) of  cluster-$n$ as the sum of two parts as
\begin{align}
{\cal H}_n&={\cal H}_n^{(0)}+{\cal H}_n^{(1)},\\
{\cal H}_n^{(0)} &=\sum_{l=1}^{n} \tilde{D}T^{z2}_l,
\\
{\cal H}_n^{(1)} &=\sum_{l=1}^{n} \frac{1}{2}J({S}^+_{l}+{S}^+_{l+1})\cdot{T}^-_{l}+{\rm h.c.}\nonumber\\
&+\sum_{l=1}^{n} J({S}^z_{l}+{S}^z_{l+1})\cdot{T}^z_{l}+\sum_{l=1}^{n+1} {D}S^{z2}_l.
\end{align}
We treat ${\cal H}_n^{(0)}$ as an unperturbed Hamiltonian and ${\cal H}_n^{(1)}$ as a perturbation Hamiltonian. 

In the present limiting case of $\De \rightarrow \infty$, the spins $\v{T}_l$ are all in the state with $T^z_l=0$. The effective Hamiltonian $\Heff^{S}(n)$ for the spins $\v{S}_l$ in cluster-$n$ is given by
\begin{align}
\Heff^{S}(0)&=DS_1^{z2}, \label{eq:heff0}\\
\Heff^{S}(1)&=-\frac{\Jeff}{2}(S^+_1S^-_{2}+S^-_1S^+_{2})+D^{\rm b}_{\rm eff}(S^{z2}_1+S^{z2}_{2})\nonumber\\
&-2\Jeff, \label{eq:heff1}\\
\Heff^{S}(n)&=-\sum_{l=1}^{n}\frac{\Jeff}{2}(S^+_lS^-_{l+1}+S^-_lS^+_{l+1})+\sum_{l=2}^{n}D_{\rm eff}S^{z2}_l\nonumber\\
&+D^{\rm b}_{\rm eff}(S^{z2}_1+S^{z2}_{n+1})-2n\Jeff \quad (n \geq 2), \label{eq:heff}
\end{align}
where
\begin{align}
\Jeff&=\frac{2J^2}{\tilde{D}},\  D_{\rm eff}=\frac{2J^2}{\tilde{D}}+D ,\  D^{\rm b}_{\rm eff}=\frac{J^2}{\tilde{D}}+D \label{eq:effpara}
\end{align}
up to the second order in $J$ and $D$.\cite{hc} 
\subsubsection{Case of ${\Do}=0$}
The effective Hamiltonians (\ref{eq:heff0}), (\ref{eq:heff1}), and (\ref{eq:heff}) respectively reduce to the form
\begin{align}
\Heff^{S}(0)&=0,\\
\Heff^{S}(1)&=-\frac{\Jeff}{2}(S^+_1S^-_{2}+S^-_1S^+_{2})\nonumber\\
&+\frac{\Jeff}{2}(S^{z2}_1+S^{z2}_{2})-2\Jeff,\\
\Heff^{S}(n)
&=\Jeff\Big[-\sum_{l=1}^{n}\frac{1}{2}(S^+_lS^-_{l+1}+S^-_lS^+_{l+1})\nonumber\\
&+\sum_{l=2}^{n}S^{z2}_l+\frac{1}{2}(S^{z2}_1+S^{z2}_{n+1})-2n\Big],  \quad (n \geq 2). 
\end{align}
The ground-state energies of $\Heff^{S}(n)$ for $n=0, 1$, and 2 are given as follows: 
\begin{align}
E_{\rm G}(1,0,{\De})&=0,\\
E_{\rm G}(3,0,{\De})&=-3\Jeff,\\
E_{\rm G}(5,0,{\De})&\simeq -5.660555382754555\Jeff.
\end{align}
The value for $n=2$ is obtained by numerical diagonalization. 
The energy per spin in the thermodynamic limit is obtained by the infinite-size DMRG method as
\begin{align}
\epsilon_{\rm G}(0,\tilde{D})\simeq -2.6995\Jeff.
\end{align}
Using these results in Eqs. (\ref{eq:lamcnnp}) and (\ref{eq:lamcninf}), we find
\begin{align}
\lambda_{\rm c}(1,0)&=3\Jeff,\\
\lambda_{\rm c}(\infty,1)&\simeq 2.399\Jeff.
\end{align}
Since $\lambda(2,1)$ estimated by Eq. (\ref{eq:lamcnnp}) with $n=2$ is less than $\lambda_{\rm c}(\infty,1)$, 
 the direct transition  takes place from the DC1 phase to the DC$\infty$ phase as long as $J$ is much smaller than $\De$. 
Rewriting $\Jeff$ by $\lambda$ and $\Delta$ using Eqs. (\ref{eq:defde}) and (\ref{eq:effpara}), we find 
\begin{align}
\lambda_{\rm c}(1,0)&\simeq{\dfrac{{2\sqrt{3}}J}{\sqrt{\Delta-1}}},\\
\lambda_{\rm c}(\infty,1)&\simeq\frac{{3.098}J}{\sqrt{\Delta-1}},
\end{align}
as long as $\Delta$ is sufficiently close to unity. Since $\Jeff/\Deff=1$, the DC$\infty$ phase belongs to the large-$D$ phase according to the phase diagram of the uniform spin-1 XXZ chain.\cite{chs} The cluster-0 is a single spin $\v{S}_l$ with no anisotropy. Hence, the DC0 phase is isotropic and paramagnetic. The ground state of  cluster-1 is given by
\begin{align}
\ket{\psi_1}=\frac{1}{\sqrt{6}}\left(\ket{\uparrow 0 \downarrow}+2\ket{000}+\ket{\downarrow 0 \uparrow}\right),
\end{align}
where $\uparrow$, 0, and  $\downarrow$ denote the state with $S^z=1, 0$, and $-1$, respectively. Since this state has $S^z_{\rm tot}(1)=0$, the DC1 phase is nonmagnetic.
\subsubsection{Case of $\Do > 0$}
Since $\Do \gg \Jeff$, we decompose the effective Hamiltonian (\ref{eq:heff}) into two parts as
\begin{align}
\Heff^{S}(n)&={\cal{H}}^{(0)}_{\rm eff}(n)+{\cal{H}}^{(1)}_{\rm eff}(n),\label{eq:heff_dev} \\
{\cal{H}}^{S(0)}_{\rm eff}(n)&=\sum_{l=2}^{n}D_{\rm eff}S^{z2}_l+D^{\rm b}_{\rm eff}(S^{z2}_1+S^{z2}_{n+1})-2n\Jeff,\label{eq:heff_dev_0}\\
{\cal{H}}^{S(1)}_{\rm eff}(n)&=-\sum_{l=1}^{n}\frac{\Jeff}{2}(S^+_lS^-_{l+1}+S^-_lS^+_{l+1}).
\end{align}
We regard ${\cal{H}}^{S(0)}_{\rm eff}(n)$ as an unperturbed Hamiltonian and ${\cal{H}}^{S(1)}_{\rm eff}(n)$ as a perturbation Hamiltonian. Although the last term of Eq. (\ref{eq:heff_dev_0}) is of $O(\Jeff)$, it is included in ${\cal{H}}^{S(0)}_{\rm eff}(n)$ for convenience. In the present case, $D_{\rm eff} \gg \Jeff >0$ and $D^{\rm b}_{\rm eff} \gg \Jeff >0$. Hence,  $S^z_l=0$ for all $l$ in the ground state of the Hamiltonian (\ref{eq:heff_dev_0}) and DC$n$ phases are nonmagnetic. Accordingly, the ground-state energy of the Hamiltonian (\ref{eq:heff_dev_0}) is given by
 \begin{align}
E_{\rm G}(2n+1,\Do,{\De})&=-2n\Jeff.
\end{align}
The perturbation Hamiltonian has no diagonal element. Hence, this estimation is valid up to the first order in $\Jeff$. From Eq. (\ref{eq:lamcnnp}), the critical values of $\lambda$ should satisfy
\begin{align}
\lambda_{\rm c}(n,n-1)&\simeq 2\Jeff.
\end{align}
Using Eq. (\ref{eq:defde}), we find
\begin{align}
\lambda_{\rm c}(n,n-1)&\simeq\frac{2\sqrt{2}J}{\sqrt{\Delta-1}}.
\end{align}
This expression does not depend on $n$. Hence, the direct transition from the DC0 phase to the DC$\infty$ phase takes place within the present accuracy. However, the intermediate DC$n$ phases might appear if higher-order corrections are included. Actually, we have numerically confirmed that the DC1 phase remains up to $\Delta \simeq 36$ for $D=4J$. Hence, it is likely to remain even for $\Delta \rightarrow \infty$. The DC$\infty$ phase is the large-$D$ phase, since $\Deff \gg \Jeff$.

\subsubsection{Case of $\Do <0$}

Since $J^2/\De \ll |\Do| $, we can assume $D_{\rm eff} \ll -|\Jeff| < 0$ and $D^{\rm b}_{\rm eff} \ll -|\Jeff|  < 0$. Hence,  $S^z_l=\pm 1$ for all $l$ in the ground state. The ground-state wave function and energy of ${\cal{H}}^{(0)}_{\rm eff}(n)$ are given by
\begin{align}
\ket{{\psi}_{n}^{(0)}} &=\ket{\updownarrow 0 \updownarrow 0...0\updownarrow}, \\
E_{\rm G}(2n+1,D,\tilde{D})&=(n-1)D_{\rm eff}+2D^{\rm b}_{\rm eff}-2nJ_{\rm eff}\nonumber\\
&=(n+1)D-\frac{2nJ^2}{\tilde{D}}.
\end{align}
Here, $\updownarrow $ stands for the state in which ${S}_l^z$ can take either $1$ or $-1$. Note that the last expression is also valid for $n=0$. Since all $S^z_l$ can take two values $\pm 1$, the ground states are $2^{n+1}$-fold-degenerate. However, the perturbation Hamiltonian  ${\cal{H}}^{(1)}_{\rm eff}(n)$ has no off-diagonal elements among these ground states, because the operators $S_l^{\pm}$ can change $S^z_l$ only by $\pm 1$. Hence, this estimation of $E_{\rm G}(2n+1,D,\tilde{D})$ is valid up to the first order in $\Jeff$. 

From Eq. (\ref{eq:lamcnnp}), the critical values of $\lambda$ should satisfy
\begin{align}
\lambda_{\rm c}(n,n-1)&\simeq \frac{2J^2}{\tilde{D}}.
\end{align}
Using Eq. (\ref{eq:defde}), we find
\begin{align}
\lambda_{\rm c}(n,n-1)&\simeq \frac{2J}{\sqrt{\Delta-1}}.
\end{align}
This expression does not depend on $n$. Hence, the direct transition from the DC0 phase to the DC$\infty$ phase takes place within the present accuracy.  However, the intermediate DC$n$ phases might appear if higher-order corrections are included.  Actually, we have numerically confirmed that the DC1 phase survives up to $\Delta \simeq 46$ for $D=-2J$. Hence, it is likely to survive even for $\Delta \rightarrow \infty$.

The $2^{n+1}$-fold degeneracy is lifted within the 3rd-order perturbation calculation in ${\cal H}_n^{(1)}$. In this order, the effective Hamiltonian between ${S}_l^z$'s  is  an antiferromagnetic Ising chain given by
\begin{align}
{\cal H}^{(n)}_{\rm Ising}&=\sum_{i=1}^{n}J^z_{\rm eff} S^z_lS^z_{l+1},
\end{align}
where $J^z_{\rm eff}=\dfrac{2J^3}{{\Do}^2}$.\cite{hc} Note that the effective spin-flip term is in the 4th order in $J$. Since $T^z_l=0$ for all $l$, the ground state of  cluster-$n$ is given by 
\begin{align}
\ket{{\psi}_{n}^{(0)}} &=\ket{\uparrow 0 \downarrow 0\uparrow...0\uparrow} 
\mbox{or} \ket{\downarrow 0 \uparrow 0\downarrow...0\downarrow} 
\end{align}
 for an even $n$. These states have $S_{\rm tot}^z(n)=\pm 1$. It is given by
\begin{align}
\ket{{\psi}_{n}^{(0)}} &=\ket{\uparrow 0 \downarrow 0\uparrow...0\downarrow} \mbox{or} \ket{\downarrow 0 \uparrow 0\downarrow...0\uparrow}
\end{align}
 for an odd $n$. These states have $S_{\rm tot}^z(n)=0$. Hence, the ground state of the DC$n$ phase is paramagnetic for an even $n$ and nonmagnetic for an odd $n$. Accordingly, the ground state corresponding to the DC$\infty$ phase in this region is the PDN-I phase.

\subsection{Limiting case of $\Delta  \gtrsim -1$}

\subsubsection{Possible ground states of cluster-$n$}

As $\Delta$ approaches $-1$, the critical values of $\lambda$  diverge as justified by the conclusion of the present subsection. From Eq. (\ref{eq:defde}), this corresponds to the limit $\De \rightarrow -\infty$.  Considering that $T_l^z$ can only take the values of $\pm 1$ in this limit, we have the following three candidates $\ket{{\psi}_{n,\alpha}^{(0)}} (\alpha=1,2,$ and $3) $ of the ground states of ${\cal H}_n^{(0)}$. The corresponding  ground-state energy is denoted by $E_{\rm G}(2n+1,{\Do},{\De};\alpha)$, including the first-order correction in ${\cal H}_n^{(1)}$.
\begin{enumerate}
\item For all $l$, ${S}_l^z=1$ and $T_l^z=-1$ or reversed.
\begin{align}
&\ket{{\psi}_{n,1}^{(0)}} 
=\ket{\uparrow\downarrow...\uparrow} \mbox{or} \ket{\downarrow\uparrow...\downarrow},\\
&E_{\rm G}(2n+1,{\Do},{\De};1) =-n|\tilde{D}|+(n+1){D}-2nJ.
\end{align}
Here, $\uparrow (\downarrow)$ stands for the state with ${S}_l^z,T_l^z =1(-1)$. Cluster-$n$ has $S^z_{\rm tot}(n)=\pm 1$ for all $n$ and the DC$n$ state is paramagnetic. The corresponding DC$\infty$ phase is the N\'eel phase.
\item  ${S}_1^z={S}_{n+1}^z=0$, otherwise, ${S}_l^z=1$ and $T_l^z=-1$  or reversed.
\begin{align}
&\ket{{\psi}_{n,2}^{(0)}} 
=\ket{0\downarrow\uparrow\downarrow ...\uparrow\downarrow 0} \mbox{or} \ket{0\uparrow\downarrow\uparrow ...\downarrow\uparrow 0},\\
&E_{\rm G}(1,{\Do},{\De};2) =0,\\
&E_{\rm G}(2n+1,{\Do},{\De};2) \nonumber\\
&=-n|\tilde{D}|+(n-1){D}-2(n-1)J, \ \ (n \geq 1).
\end{align}
Here, $0$ stands for the state with ${S}_l^z=0$.  Cluster-$0$ has $S^z_{\rm tot}(0)=0$, but cluster-$n$ with $n \geq 1$ has $S^z_{\rm tot}(n)=\pm 1$. Hence, the DC$0$ phase is nonmagnetic and the DC$n$ phases with  $n \geq 1$ are paramagnetic. The corresponding DC$\infty$ phase is the N\'eel phase.
\item For all $l$, ${S}_l^z=0$ and $T_l^z=\pm 1$.  The ground-state wave function and energy are given by
\begin{align}
&\ket{{\psi}_{n,3}^{(0)}} =\ket{0\updownarrow 0 \updownarrow 0...0\updownarrow 0}, \\
&E_{\rm G}(2n+1,{\Do},{\De}; 3)=-n|\tilde{D}|-2n\frac{J^2}{|\De|},\label{eq:eg3}
\end{align}
including the second-order correction in $J$ for later convenience. Here, $\updownarrow $ stands for the state in which ${T}_l^z$ can take either $1$ or $-1$. 

Up to this accuracy, the ground state is $2^{n}$-fold degenerate, because each $T_{l}^z$ can take values of $\pm 1$. The degeneracy is lifted within the 3rd order perturbation calculation in ${\cal H}_n^{(1)}$. In this order, the effective Hamiltonian between ${T}_l^z$'s  is  an antiferromagnetic Ising chain given by
\begin{align}
{\cal H}^{(n)}_{\rm Ising}&=\sum_{i=1}^{n-1}J^z_{\rm eff} T^z_lT^z_{l+1},
\end{align}
where $J^z_{\rm eff}=\dfrac{2J^3}{{\De}^2}$.\cite{hc} Note that the effective spin-flip term is in the 4th order in $J$.  Since $S^z_l=0$ for all $l$, the ground state of  cluster-$n$ is given by
\begin{align}
\ket{{\psi}_{n,3}^{(0)}} &=\ket{0\uparrow 0 \downarrow 0\uparrow...0} \mbox{or} \ket{0\downarrow 0 \uparrow 0\downarrow...0}.
\end{align}
Hence, the ground state of cluster-$n$ has $S^z_{\rm tot}(n)=\pm 1$ for an odd $n$ and  $S^z_{\rm tot}(n)=0$ for an even $n$. The corresponding DC$\infty$ phase in this region is the PDN-II phase.

\end{enumerate}

\subsubsection{Case of $D <J$}
The ground state is $\ket{{\psi}_{n,1}^{(0)}}$ for all $n$. From Eq. (\ref{eq:lamcnnp}), we obtain
\begin{align}
\lambda_{\rm c}(n,n-1)&\simeq |\tilde{D}|+2J.
\end{align}
Using Eq. (\ref{eq:defde}), this leads to 
\begin{align}
{\lambda_{\rm c}(n,n-1)}&\simeq\frac{4J}{1+\Delta}. \label{eq:lambdac_2}
\end{align}
This expression tends to $\infty$ as $\Delta\rightarrow -1-0$ as assumed. It  does not depend on $n$. Hence, the direct transition from the paramagnetic DC0 phase to the N\'eel-ordered DC$\infty$ phase takes place within the present accuracy.  However, the intermediate DC$n$ phases might appear if higher-order corrections are included.  Actually, we have numerically confirmed that the DC1 phase survives up to $\lambda \simeq 132J$ $ (\Delta \simeq -0.97)$ for $D=-2J$ and $D=0$. Hence, it is likely to survive even for $\lambda \rightarrow \infty$ ($\Delta  \rightarrow -1$). 
\subsubsection{Case of $J<D <2J$}

The ground state is $\ket{{\psi}_{n,2}^{(0)}}$ for all $n$. 
From Eq. (\ref{eq:lamcgen}), the transition point from the DC0 phase to the  DC$n$ phase is given by
\begin{align}
\lambda_{\rm c}(n,0)&=|\tilde{D}|-\frac{(n-1)}{n}({D}-2J).
\end{align}
Using Eq. (\ref{eq:defde}), this implies that
\begin{align}
{\lambda_{\rm c}}(n,0)&=\left(1-\frac{1}{n}\right)\frac{2}{(1+\Delta)}(2J-{D}).
\end{align}
Since this is an increasing function of $n$, the direct transition from the nonmagnetic DC0 phase to the N\'eel-ordered DC$\infty$ phase takes place at
\begin{align}
{\lambda_c}(\infty,0)&=\frac{2}{(1+\Delta)}(2J-{D}).
\end{align}
Since $2J >D$,  $\lambda_{\rm c}$ tends to $\infty$ as $\Delta\rightarrow -1-0$, as assumed. 
Note that this contribution vanishes at $D=2J$.

\subsubsection{Case of $2J \leq D $}
The ground state is $\ket{{\psi}_{n,3}^{(0)}}$ for all $n$. 
Using Eqs. (\ref{eq:lamcnnp}) and (\ref{eq:eg3}),  we have
\begin{align}
\lambda_{\rm c}(n,n-1)
&=|\De|+2\frac{J^2}{|\De|
}.
\end{align}
Using Eq. (\ref{eq:defde}), we find the critical value of $\lambda$ as
\begin{align}
\lambda_{\rm c}(n,n-1)&\simeq\frac{2\sqrt{2}J}{\sqrt{1-\Delta^2}}.\label{eq:lambdac_3}
\end{align}
This expression tends to $\infty$ as $\Delta\rightarrow -1-0$, as assumed. Since this does not depend on $n$, the direct transition from the nonmagnetic DC0 phase to the PDN-II DC$\infty$ phase takes place within the present accuracy. 
 Actually, we have numerically confirmed that the DC1 phase does not appear for $\lambda \gtrsim 0.62J$ $(\Delta \lesssim 10.8)$ for $D=4J$. 

\subsection{Limit of $\Do \rightarrow -\infty$ 
}\label{sect:negD}

\subsubsection{Hamiltonian}
We express the Hamiltonian (\ref{ham3}) of cluster-$n$ as the sum of two parts as
\begin{align}
{\cal H}_n&={\cal H}_n^{(0)}+{\cal H}_n^{(1)},\\
{\cal H}_n^{(0)} &=\sum_{l=1}^{n+1} {D}S^{z2}_l,
\\
{\cal H}_n^{(1)} &=\sum_{l=1}^{n} \frac{1}{2}J({S}^+_{l}+{S}^+_{l+1})\cdot{T}^-_{l}+{\rm h.c.}\nonumber\\
&+\sum_{l=1}^{n} J({S}^z_{l}+{S}^z_{l+1})\cdot{T}^z_{l}+\sum_{l=1}^{n} \tilde{D}T^{z2}_l.
\end{align}
We treat ${\cal H}_n^{(0)}$ as an unperturbed Hamiltonian and ${\cal H}_n^{(1)}$ as a perturbation Hamiltonian. 

\subsubsection{$\dfrac{\Delta-1}{2}\lambda > 2J$}
The ground-state energy of cluster-$n$  is given by
\begin{align}
E_{\rm G}(2n+1,\Do,\De)&=-(n+1)|\Do|.
\end{align}
 The phase boundaries are given by
\begin{align}
\lambda_{\rm c}(n,n-1)&=0
\end{align}
using Eqs. (\ref{eq:lamcnnp}) and (\ref{eq:lamcninf}). 
Hence, the ground state is always the DC0 phase in this region. 
\subsubsection{$\dfrac{\Delta-1}{2}\lambda < 2J$}

\begin{figure}
\centerline{\includegraphics[height=5.5cm]{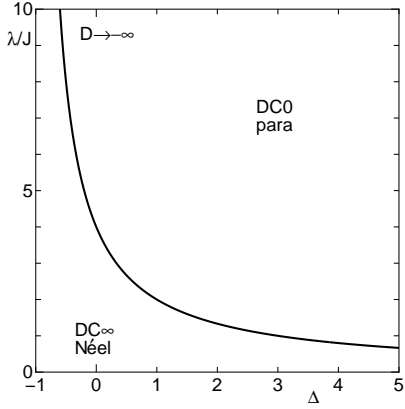}}
\caption{ Phase diagram for ${\Do}\rightarrow -\infty$. }
\label{fig:dominf}
\end{figure}

In this case, all the spins $T_l^z$ are $\pm 1$. Since the first term of ${\cal H}_n^{(1)}$ cannot change $T_l^z$ by $\pm 2$, this spin-flip term does not play a role up to the second order in  ${\cal H}_n^{(1)}$. Hence, the quantum fluctuation is suppressed resulting in the N\'eel-ordered ground state. 
 The ground-state energy of cluster-$n$ and the ground-state energy per unit cell of the DC$\infty$ phase  are given by
\begin{align}
E_{\rm G}(2n+1,\Do,\De)&=-(n+1)|\Do|+n\left(\dfrac{\Delta-1}{2}\lambda-2J\right),\\
\epsilon_{\rm G}(D,\tilde{D};\mbox{N\'eel})&=-|\Do|+\dfrac{\Delta-1}{2}\lambda-2J.
\end{align}
The phase boundary satisfies
\begin{align}
\lambda_c(n,n-1)&=\dfrac{4J}{\Delta+1}\label{eq:lamd_0_neel}
\end{align}
from Eqs. (\ref{eq:lamcnnp}) and (\ref{eq:lamcninf}). 
This expression does not depend on $n$. Hence, the direct transition from the DC0 phase to the DC$\infty$ phase takes place at this point 
within the present accuracy.  However, the intermediate DC$n$ phases might appear if higher-order corrections are included. 
 Cluster-0 is a single spin $\v{S}_i$ that can take the states with $S^z_i=\pm 1$ because of the negative $\Do$. Hence, the DC0 phase is paramagnetic. Note that the critical value of $\lambda$ given by Eq. (\ref{eq:lamd_0_neel}) satisfies the applicability condition of the present estimation as
\begin{align}
\dfrac{\Delta-1}{2}\lambda=\dfrac{2J(\Delta-1)}{\Delta+1}<2J.
\end{align} 
The phase diagram is shown in Fig. \ref{fig:dominf}.

\subsection{Limit of $\Do \rightarrow \infty$ 
}\label{sect:posD}

In this case, the state of $S$-spins is limited to $S^z_l=0$. The effective Hamiltonian for the spins $\v{T}_l$ is given by the spin-1 XY chain with single-site anisotropy
\begin{align}
\Heff^{T}(n)&=-\sum_{l=1}^{n-1}\frac{\tJeff}{2}(T^+_lT^-_{l+1}+T^-_lT^+_{l+1})\nonumber\\
&+\sum_{l=1}^{n}\tDeff T^{z2}_l-2n\tJeff, \label{eq:Heff}\\
\tJeff&=\frac{2J^2}{{D}},\  \tDeff=\frac{2J^2}{{D}}+\De \label{eq:Heffpara} 
\end{align}
up to the second order in $J$.

\begin{enumerate}
\item Boundary between DC$n$ phases

The effective Hamiltonians of  cluster-$n$ are given by
\begin{align}
\Heff^{T}(0)&=0,\\
\Heff^{T}(1)&=\tDeff T^{z2}_1-{2}\tJeff
\end{align}
for $n=0$ and 1. The eigenvalues are given by
\begin{align}
E_{\rm G}(1,\Do,\De)&= 0,\\
E_{\rm G}(3,\Do,\De)&=\left\{
\begin{array}{ll}
\tDeff-2\tJeff& \tDeff <0,\\
 -2\tJeff&   \tDeff >0.
\end{array}
\right.
\end{align}
Using these values, we find
\begin{align}
\lambda_{\rm c}(1,0)&=\left\{
\begin{array}{ll}
\tDeff-2\tJeff& \tDeff <0,\\
 -2\tJeff&   \tDeff >0,
\end{array}
\right.\\
\lambda_{\rm c}(\infty,0)&=-\epsilon_{\rm G}(D,\tilde{D}).
\end{align}
\begin{figure}
\centerline{\includegraphics[height=5.5cm]{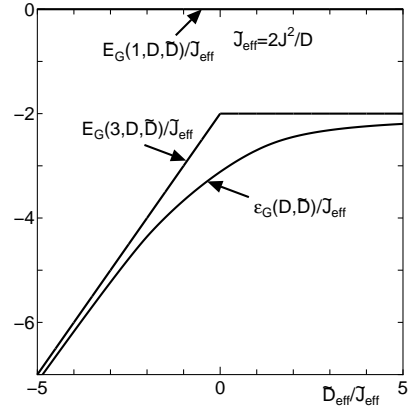}}
\caption{Ground-state energy  per unit cell $\epsilon_{\rm G}(\Do,\De)$ of $\Heff^{T}(n)$ with $n\rightarrow\infty$. The ground-state energies $E_{\rm G}(2n+1,\Do,\De)$ of $\Heff^{T}(n)$ with $n=0$ and 1 are also shown.}
\label{fig:largeD_gsene}
\end{figure}
According to the numerical values of $\epsilon_{\rm G}(D,\tilde{D})$ estimated by the DMRG calculation for the effective Hamiltonian (\ref{eq:Heff}) and $E_{\rm G}(3,\Do,\De)$  shown in  Fig. \ref{fig:largeD_gsene}, we find $\lambda_{\rm c}(\infty,0) > \lambda_{\rm c}(1,0)$ for all values of $\tilde{J}_{\rm eff}$ and $\tilde{D}_{\rm eff}$. Hence, the direct transition from the DC0 phase to the DC$\infty$ phase takes place at $\lambda=\lambda_{\rm c}(\infty,0)$  as long as $J$ is much smaller than $\De$.

\item DC$\infty$ phases

According to Ref. \citen{chs}, the ground state of the XY model (\ref{eq:Heff}) for $n \rightarrow \infty$ is the large-$D$ phase for $0.35\tJeff \le \tDeff$. It is on the critical line between the conventional XY phase and the Haldane phase for $0.35\tJeff \ge \tDeff \ge -2\tJeff$, and it is on that between the XY-II  and N\'eel phases for $\tDeff < -2\tJeff$. Considering the continuity to the numerically obtained phase diagram, the ground state of the original Hamiltonian (\ref{ham3})  for a large $D$ is actually the Haldane or N\'eel phase owing to the Ising component of the effective exchange interaction that appears as a higher-order correction in $J$, as discussed in Ref. \citen{hc}. The N\'eel phase of the effective Hamiltonian (\ref{eq:Heff}) corresponds to the PDN-II phase for the original Hamiltonian (\ref{hama}). Hence, the following two phase boundaries are present within the DC$\infty$ phase.
\begin{enumerate}
\item Haldane-PDN-II phase boundary

From the investigation of the spin-1 XXZ chain, the critical point is known to be given by $\tDeff \simeq -2\tJeff$.\cite{chs} Using  Eqs. (\ref{eq:defde}) and (\ref{eq:Heffpara}), 
the phase boundary between the Haldane phase and the PDN-II phase within the DC$\infty$ ground state is given by
\begin{align}
\lambda_{\rm c}({\rm H, PDN\mbox{-}II})&\simeq \frac{6}{1-\Delta}\tJeff.
\end{align}
\item Haldane-Large-$D$ phase boundary

From the investigation of the spin-1 XXZ chain, the critical point is known to be given by
$\tDeff \simeq 0.35\tJeff$.\cite{chs}  Using  Eqs. (\ref{eq:defde}) and (\ref{eq:Heffpara}), the boundary between the Haldane  and large-$D$ phases within the DC$\infty$ ground state is given by
\begin{align}
\lambda_{\rm c}({\rm H,LD})&\simeq \frac{1.3}{1-\Delta}\tJeff.
\end{align}
\end{enumerate}
\end{enumerate}

\begin{figure}
\centerline{\includegraphics[height=5.5cm]{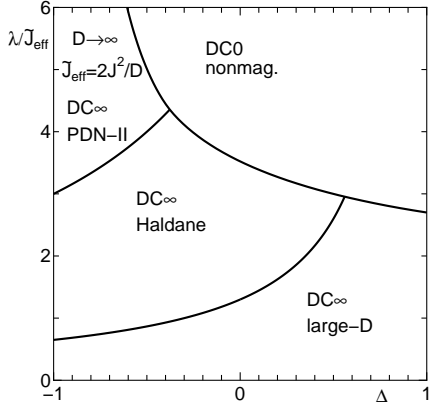}}
\caption{ Phase diagram for ${\Do}\rightarrow \infty$. }
\label{fig:dopinf}
\end{figure}

The phase diagram is summarized in Fig. \ref{fig:dopinf}.
\section{Low-Temperature Behavior of Magnetic Susceptibility and  Entropy}

To characterize the DC$n$ phases using measurable physical quantities, we examine the low-temperature behaviors of the longitudinal magnetic susceptibility $\chi$ and the entropy ${\mathcal S}$. They can be determined by the method described in Refs. \citen{hts} and \citen{htanis}. In the paramagnetic phases, the total spin of each cluster-$n$ takes two values, i.e., $\szcl{n}=\pm 1$, except in the DC0 phase with $\Do=0$. Hence, the susceptibility of each cluster-$n$ is $1/T$. Since the length of  cluster-$n$ is $n+1$, the magnetic susceptibility of the total chain behaves as
\begin{align}
\chi \simeq \frac{L}{(n+1)T},
\end{align}
at low temperatures. The entropy behaves as 
\begin{align}
{\mathcal{S}} \simeq \dfrac{L\ln 2}{n+1},
\end{align}
in the paramagnetic DC$n$ phase, since each cluster-$n$ can take two states. 

In the DC0 phase with $\Do=0$, the total spin of each cluster-$n$ can take  $\szcl{n}=\pm 1$ 
and 0. Hence, the susceptibility of each cluster-$n$ is $3/(2T)$, and  the magnetic susceptibility of the total chain behaves as
\begin{align}
\chi \simeq \frac{3L}{2(n+1)T},
\end{align}
at low temperatures. The entropy behaves as 
\begin{align}
{\mathcal{S}} \simeq \dfrac{L\ln 3}{n+1},
\end{align}
since each cluster-$n$ can take three states. 

For the nonmagnetic DC$n$ phase, $\chi$ and ${\mathcal S}$ tend to 0 at low temperatures. Hence, it is difficult to distinguish DC$n$ phases with different $n$ values from these quantities. However,  the residual entropy remains at the ground-state phase boundary. 
On the phase boundary between the DC$(n-1)$ and DC$n$ phases, two types of cluster, i.e., cluster-$(n-1)$ and cluster-$n$, coexist. 
Denoting the number of cluster-$n$'s by $N_n$, 
the total number of states of the entire chain is given by 
\begin{align}
W&=\frac{(N_{n-1}+N_{n})!}{N_{n-1}!N_{n}!}g_{n-1}^{N_{n-1}}g_n^{N_{n}}\nonumber\\
&\simeq\frac{(N_{n-1}+N_{n})^{N_{n-1}+N_{n}}}{N_{n-1}^{N_{n-1}}N_{n}^{N_{n}}}g_{n-1}^{N_{n-1}}g_n^{N_{n}},
\label{comb}
\end{align}
in the thermodynamic limit. Here, $g_n$ is the ground-state degeneracy of  cluster-$n$. In the nonmagnetic DC$n$ phases, $g_n=1$, in the paramagnetic DC0 phase with $\Do=0$, $g_n=3$, and in other paramagnetic DC$n$ phases, $g_n=2$. 
Denoting the total number of clusters by $\Nc =N_{n-1}+N_{n}$, the fraction of cluster-$(n-1)$ is given by $x=N_{n-1}/\Nc$. Then, Eq.~(\ref{comb}) is rewritten as 
\begin{align}
W&=\frac{1}{x^{\Nc x}(1-x)^{\Nc (1-x)}}g_{n-1}^{\Nc x}g_{n}^{\Nc(1-x)}. 
\end{align}
 The length of the chain is expressed as $L=nN_{n-1}+(n+1)N_{n}$.\cite{mistype}  Hence, we have $\Nc=L({n+1-x})^{-1}$. 
 Using these notations, the entropy ${\cal S}$ is expressed as 
\begin{align}
{\cal S}&=\ln W 
=-\frac{L}{n+1-x}\left[x\ln x +(1-x)\ln(1-x)-(1-x){\ln g_n}-x{\ln g_{n-1}}\right].
\label{entrop}
\end{align}
Optimizing ${\cal S}$ with respect to $x$, we have
\begin{align}
\frac{\partial {\cal S}}{\partial x}&=
-\frac{L}{(n+1-x)^2}\left[x\ln x +(1-x)\ln(1-x)-(1-x){\ln g_n}-x{\ln g_{n-1}}\right]\nonumber\\
&-\frac{L}{n+1-x}\left(\ln \frac{x}{1-x}+\ln \frac{g_n}{g_{n-1}}\right)=0.\label{eq:ent_opt_0}
\end{align}
This is simplified as
\begin{align}
n\ln \frac{(1-x)}{g_n}&=(n+1)\ln \frac{x}{g_{n-1}}.\label{eq:ent_opt}
\end{align}
Using Eq. (\ref{eq:ent_opt_0}), the expression for the entropy (\ref{entrop}) is also simplified as
\begin{align}
{\cal S}&={L}\left(\ln \frac{x}{1-x}+\ln \frac{g_n}{g_{n-1}}\right).
\end{align}
The numerical solution of Eq. (\ref{eq:ent_opt}) is used to calculate the residual entropy as tabulated in Table \ref{tab:entro_res} for small $n$ values. 
\begin{table}
\begin{tabular}{ccccc}
\hline
    $n-1$  &   $n$  &   $g_{n-1}$  &   $g_n$  &     ${\cal S}/L$ \\
\hline\hline
    0  &   1  &   1  &   1  &     0.4812118 \\
    0  &   1  &   1  &   2  &     0.6931472 \\
    0  &   1  &   1  &   3  &     0.8341152 \\
    0  &   1  &   2  &   1  &     0.8813736 \\
    0  &   1  &   2  &   2  &     1.0050525 \\
    0  &   1  &   3  &   1  &     1.1947632 \\
    0  &   1  &   3  &   3  &     1.3327058 \\
\hline
    1  &   2  &   1  &   1  &     0.2811996 \\
    1  &   2  &   1  &   2  &     0.4196176 \\
    1  &   2  &   2  &   1  &     0.4812118 \\
    1  &   2  &   2  &   2  &     0.5705797 \\
\hline
    2  &   3  &   1  &   1  &     0.1994606 \\
    2  &   3  &   1  &   2  &     0.3024795 \\
    2  &   3  &   2  &   1  &     0.3331360 \\
    2  &   3  &   2  &   2  &     0.4018119 \\
\hline
\end{tabular}
\caption{Residual entropies at the DC$(n-1)$-DC$n$ phase boundaries for $n \leq 3$.}\label{tab:entro_res}
\end{table}

\section{Correspondence between the Anisotropic Rung-Alternating Ladder and the Anisotropic Mixed Diamond Chain}

\begin{figure}
\centerline{\includegraphics[width=5cm]{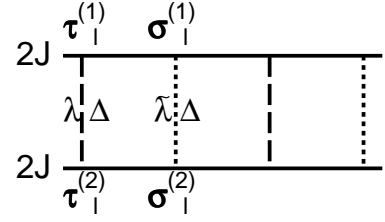}}
\caption{ Structure of the anisotropic rung-alternating ladder. }
\label{fig:ral}
\end{figure}

The AMDC  (\ref{hama}) can be regarded as a limiting case of the ARAL whose  Hamiltonian is given by
\begin{align}
&{\cal{H}}_{\rm ARAL}=\sum_{l=1}^{L}\sum_{\alpha=1}^{2}2J({\v{\sigma}}^{(\alpha)}_{l}{\v{\tau}}^{(\alpha)}_{l}+{\v{\sigma}}^{(\alpha)}_{l}{\v{\tau}}^{(\alpha)}_{l+1})\nonumber\\
&+\sum_{l=1}^{L}\left[\tilde{\lambda}\left(\Delta{{\sigma}}^{(1)z}_{l}{{\sigma}}^{(2)z}_{l}+({{\sigma}}^{(1)x}_{l}{{\sigma}}^{(2)x}_{l}+{{\sigma}}^{(1)y}_{l}{{\sigma}}^{(2)y}_{l})\right)\right.\nonumber\\
&\left.+\lambda\left(\Delta{\tau}^{(1)z}_{l}{\tau}^{(2)z}_{l}+({\tau}^{(1)x}_{l}{\tau}^{(2)x}_{l}+{\tau}^{(1)y}_{l}{\tau}^{(2)y}_{l})\right)\right],
\end{align}
where the notations are chosen to maintain correspondence with our Hamiltonian  (\ref{hama}) of the AMDC. The lattice structure is shown in Fig. \ref{fig:ral}. The operators ${\v{\sigma}}^{(\alpha)}_{l}$ and ${\v{\tau}}^{(\alpha)}_{l} (\alpha=1,2)$ are spin-1/2 operators.

Tonegawa {\it et al.}\cite{tone_prv} have investigated the ground-state phase diagram of the ARAL. They found large-$D$, Haldane, N\'eel, and triplet-dimer-singlet-dimer (TD-SD) phases for $\Delta <1$, and N\'eel, Haldane, and ferromagnetic-singlet-dimer (F-SD) phases for $\Delta >1$. In the limit $\tilde{\lambda}\rightarrow -\infty$, this model tends to the AMDC (\ref{hama}) by identifying $\v{S}_l\equiv\v{\sigma}_{l,1}+\v{\sigma}_{l,2}$.  In this limit, the single-ion anisotropy $\Do$ on the spin $\v{S}_l$ is given by
\begin{align}
\Do&=\dfrac{\tilde{\lambda}(\Delta-1)}{2}. \label{eq:raldo}
\end{align}
This tends to $\infty (-\infty)$ for $\Delta < ( > ) 1$ as $\tilde{\lambda}\rightarrow -\infty$. Namely, the limit of a strong $\tilde{\lambda}$ corresponds to the strong anisotropy limit $D \rightarrow \pm \infty$ discussed in sects. \ref{sect:negD} and \ref{sect:posD}, unless $\Delta$ is so close to unity that $|\Delta -1| \lesssim O(J/\tilde{\lambda})$.
\begin{figure}
\centerline{\includegraphics[height=5.5cm]{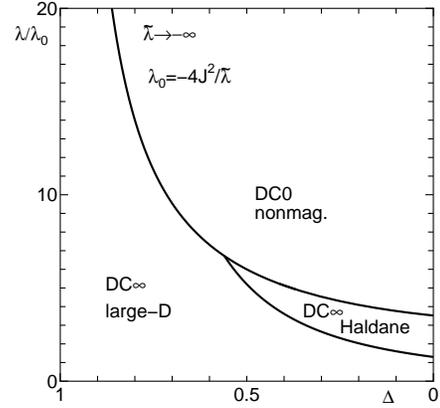}}
\caption{ Ground-state phase diagram of the ARAL for $\tilde{\lambda}\rightarrow -\infty$ with $0 < \Delta < 1$, corresponding to  the AMDC with ${\Do}\rightarrow \infty$. }
\label{fig:dopinfv}
\end{figure}
\begin{figure}
\centerline{\includegraphics[height=5.5cm]{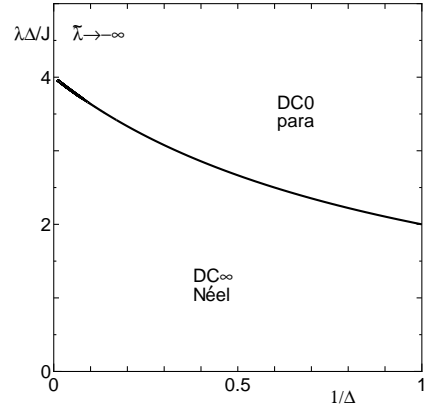}}
\caption{ Ground-state phase diagram of the ARAL for  $\tilde{\lambda}\rightarrow -\infty$ with $\Delta > 1$, corresponding to the AMDC with ${\Do}\rightarrow -\infty$. To compare it with the phase diagram of the ARAL obtained by Tonegawa {\it et al.},\cite{tone_prv} the results are plotted on the $1/\Delta$-$\lambda\Delta$ plane. }
\label{fig:dominfv}
\end{figure}
\begin{figure}
\centerline{\includegraphics[height=5.5cm]{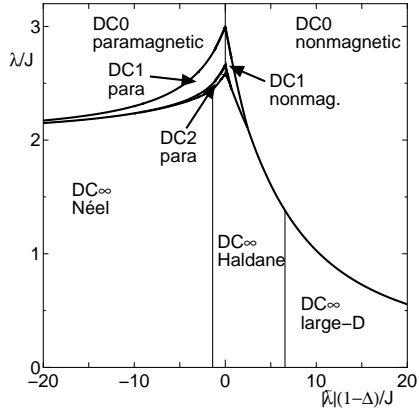}}
\caption{ Phase diagram of the ARAL for $\Delta \simeq 1$ with $\tilde{\lambda} \ll -J < 0 $.}
\label{fig:almost_isotropic}
\end{figure}
Our results for the phase diagram are compared with those for the ARAL by Tonegawa {\it et al.}\cite{tone_prv} in the following way.

In the region $0 \leq \Delta < 1$, the phase diagram of  Fig. \ref{fig:dopinf} is replotted in Fig. \ref{fig:dopinfv} taking the vertical axis as ${\lambda}/{\lambda_0}$ where
\begin{align}
{\lambda_0}\equiv -\frac{4J^2}{\tilde{\lambda}}.
\end{align} 
Defining $x={\tDeff}/{\tJeff}$ and using Eqs. (\ref{eq:defde}) and  (\ref{eq:Heffpara}), we find  
\begin{align}
\frac{\lambda(\Delta-1)}{2}=\tJeff(x-1).\label{eq:Descale}
\end{align}
On the DC0-DC$\infty$ phase boundary, substituting  $\lambda=\lambda_{\rm c}(\infty,0)=-\epsilon_{\rm G}(D,\tilde{D})$ to Eq. (\ref{eq:Descale}) and rescaling as $\epsilon_{\rm G}(D,\tilde{D})=\tJeff f(x)$, we obtain
\begin{align}
\Delta=1-\frac{2\left(x-1\right)}{f(x)}.\label{eq:Delta}
\end{align}
On the other hand, we have
\begin{align}
\frac{\lambda_{\rm c}(\infty,0)}{\lambda_0}=-\frac{\tJeff}{\lambda_0} f(x)=\frac{f(x)}{\Delta-1}=\frac{f(x)^2}{2\left(1-x\right)}.\label{eq:lamc}
\end{align}
The function $f(x)$ is numerically determined as a function of $x$ by the infinite-size DMRG method. The DC0-DC$\infty$ phase boundary  in Fig. \ref{fig:dopinfv} is drawn using Eqs. (\ref{eq:Delta}) and (\ref{eq:lamc}).

The nonmagnetic DC0 phase corresponds to the TD-SD phase of the ARAL. In the ARAL, the N\'eel phase with an antiferromagnetic correlation on the rungs appears between the TD-SD and large-$D$ phases. This phase disappears in the AMDC owing to the strong XY-like ferromagnetic rung coupling $\tilde{\lambda}$. In this limit, the state of the pair $\v{\sigma}_l^{(1)}$ and $\v{\sigma}_l^{(2)}$ becomes the entangled state $\frac{1}{\sqrt{2}}(\ket{\uparrow\downarrow}+\ket{\uparrow\downarrow})$ and the composite spins $\v{T}_l$ are decoupled up to the lowest order in $J$.  The effective coupling between the $\v{T}_l$ spins leads to the Haldane ground state for $\Delta \gtrsim 0.56$. The Haldane phase around the isotropic point $\Delta=1$ in the ARAL shrinks to the region $0 < 1- \Delta < O(J/\tilde{\lambda})$ and  disappears in the phase diagram in Fig. \ref{fig:dopinfv}

In the region $\Delta > 1$, the paramagnetic DC0 phase corresponds to the F-SD phase of the ARAL. However, the spin pairs on the neighboring ferromagnetic rungs, which correspond to the neighboring spins $\v{S}_l$ in the AMDC, are antiferromagnetically coupled with each other in the ARAL owing to the effective interaction between  the spins $\v{S}_l$, which stems from the finite $\tilde{\lambda}$. The Haldane phase that appears between the F-SD and N\'eel phases in the ARAL disappears in the AMDC, since there is no room for quantum fluctuation in this regime, as described in sect. \ref{sect:negD}.
   The Haldane phase around the isotropic point $\Delta=1$ in the ARAL also shrinks to the region $0 < \Delta-1 < O(J/\tilde{\lambda})$ and  disappears in the phase diagram in Fig. \ref{fig:dominfv}.

For $|\Delta-1| \sim O(J/\tilde{\lambda})$, the ARAL corresponds to the AMDC with anisotropy $\Do$ given by Eq. (\ref{eq:raldo}) with negligible anisotropy on the spins  ${\v{\tau}}^{(\alpha)}_{l}$. This model has been investigated in Ref. \citen{htanis}. A corresponding phase diagram is redrawn in terms of the parameters of the ARAL in Fig. \ref{fig:almost_isotropic}, using the data of  Fig. 2(a) of Ref. \citen{htanis}. This explains the presence of the Haldane phase of the ARAL at around $\Delta \simeq 1$.

\section{Summary and Discussion}

We examined the AMDC with spins 1 and 1/2, which has the single-site anisotropy $D$ on 
  spin-1 sites and exchange anisotropy $\Delta$ on the bond connecting spin-1/2 sites. 
In the ground-state phase diagram, there are  DC$n$ phases with a finite $n$   if $\lambda$ is above the critical value, which depends on $D$ and $\Delta$.  
As in the isotropic case, the DC$n$ ground state with a finite $n$ is an alternating array of finite-length cluster-$n$'s and singlet dimers. The low-temperature behaviors of magnetic susceptibility and entropy, which characterize the DC$n$ phases and phase boundaries between them, are investigated. 
For a smaller $\lambda$, the ground-state phase is one of the five DC$\infty$ phases, i.e., the N\'eel,  Haldane, large-$D$, and two different period-doubled N\'eel phases.  Various limiting cases are analytically discussed.  The relationship of the present model with the anisotropic rung-alternating ladder is also discussed.

The DC$\infty$ phases correspond to the ground states of the spin-1 chain with alternating single-site anisotropies. Actually, in the latter model, the alternation of two anisotropies is rather artificial. 
 In the AMDC, the two anisotropies concerns the different spins $\v{S}_l$ and $\v{\tau_{l}^{(\alpha)}} (\alpha=1,2)$. Hence, it is natural that $\Do$ and $\De$ are different.

If the corresponding material is synthesized, each phase and/or phase boundary can be distinguished by the low-temperature magnetic susceptibility and residual entropy. 
The DC$n$ phases with $n \ge 2$ are found to be extremely narrow. Hence, it would be difficult to identify them experimentally. 
 Actually, the width of the DC$n$ phases with a higher $n$ is largest around the isotropic point $D=0$ and $\Delta=1$. Namely, the anisotropy suppresses these phases. On the other hand, the DC$\infty$ PDN phases are only realized if the two types of anisotropy coexist. 

The author thanks K. Takano, T. Tonegawa, and K. Okamoto for valuable comments and discussions. The numerical diagonalization program is based on the package TITPACK ver. 2 coded by H. Nishimori.  Part of the numerical computation in this work was carried out using the facilities of the Supercomputer Center, Institute for Solid State Physics, University of Tokyo,  and the  Yukawa Institute Computer Facility, Kyoto University.  This work is  supported by a Grant-in-Aid for Scientific Research (C) (25400389) from the Japan Society for the Promotion of Science.

\appendix


\begin{thebibliography}{50}
\bibitem{mistype} The corresponding equation was mistyped in Ref. \citen{hts}.
\end{thebibliography}
\end{document}